\pdfoutput=1
\documentclass[11pt]{article}
\usepackage[a4paper,margin=1in]{geometry}
\usepackage[T1]{fontenc}
\usepackage[utf8]{inputenc}
\usepackage{lmodern}
\usepackage{microtype}
\usepackage{amsmath,amssymb}
\usepackage{graphicx}
\usepackage{booktabs}
\usepackage{xcolor}\definecolor{ykmark}{HTML}{007F3F}
\usepackage[font=small,labelfont=bf]{caption}
\usepackage{enumitem}\setlist{nosep,leftmargin=1.5em}
\usepackage[numbers,sort&compress]{natbib}
\usepackage[colorlinks=true,linkcolor=blue!50!black,citecolor=blue!50!black,urlcolor=blue!50!black]{hyperref}
\usepackage{url}
\usepackage{orcidlink}
\usepackage{placeins}
\usepackage{adjustbox}
\usepackage{float}
\newfloat{algorithm}{tbp}{loa}
\floatname{algorithm}{Algorithm}
\usepackage{listings}
\title{\vspace{-0.4cm}When Does Fine-Tuning Hurt? Cross-Domain Generalization in Low-Resource Cantonese ASR}
\author{Shun Cheung\textsuperscript{1}, Victor Lam\textsuperscript{2}, and Xihui Wu\textsuperscript{2}}
\date{}

\begin{document}
{\let\newpage\relax\maketitle}
\vspace{-1.2cm}
\begin{center}\small \textsuperscript{1}\,The Chinese University of Hong Kong, Hong Kong SAR, China\\ \textsuperscript{2}\,Dark Unit AI Limited\\ \end{center}

\begin{abstract}When fine-tuning automatic speech recognition (ASR) models on 6.3 hours of scripted Cantonese speech from the Cantonese Life Scenarios corpus, adaptation outcomes differ across backbones with different zero-shot capabilities. For Whisper-small, whose zero-shot Cantonese accuracy is weak, we developed our competition submission W500 using source-conditional two-stage fine-tuning: alternating steps draw from a single source so external WenetSpeech-Yue audio updates only the encoder while competition utterances update the full model. On the competition platform's 200-utterance blind hidden test set, W500 achieved an official score of 69.49 (24.73\% CER). Under unified symmetric scoring, it reached 8.09\% CER on the 1,900-utterance public in-domain block (which also served as a Stage 2 eligibility filter) and reduced CER on three independent spontaneous conversations (MagicData-Conv) from 62.69\% zero-shot to 25.04\%. In a twelve-run Whisper capacity matrix, validation and the in-domain test set rank LoRA and full fine-tuning differently, and scaling beyond medium yields minimal validation gains. Traditional-versus-simplified script mismatch drives unnormalized out-of-domain errors, and a matched three-seed control with MDCC external audio shows that freezing the decoder on external steps costs validation accuracy and does not reduce conversational CER relative to a conventional mixed-batch recipe at equal external exposure. In contrast, fine-tuning Qwen3-ASR, which already recognizes Cantonese well zero-shot, with the ChinaVoices LoRA recipe on the same scripted speech lowers 1.7B in-domain CER from 9.69\% to 3.31\% while raising MagicData-Conv CER from 9.09\% to 14.12\%. Lowering the learning rate tenfold to $10^{-5}$ preserves the in-domain gain and reduces the conversational degradation by 47\% at 1.7B and 78\% at 0.6B. A cross-hardware replication of all seventeen Qwen3-ASR runs plus the public WSYue-ASR-eval benchmark reproduces both the degradation and the monotonic learning-rate ordering, and the excess conversational error persists after folding numerals and written variants. Code and the W500 checkpoint are public; per-utterance predictions are available for checking.\end{abstract}

\section{Introduction}\label{sec:intro}

The AI Dimsum Cup preliminary Cantonese ASR task \citep{dimsum} supplied about six hours of scripted Cantonese speech from the Cantonese Life Scenarios corpus \citep{lifescen} and evaluated submissions on an in-distribution hidden test set. Adapting ASR backbones in this low-resource regime depends on their zero-shot starting point: models with weak zero-shot Cantonese recognition need external audio to expand acoustic coverage without leaking external transcription conventions into the decoder. Models that already recognize Cantonese well zero-shot instead risk losing generalization on spontaneous speech and public benchmarks after scripted fine-tuning. For a weak-prior backbone, we built and released W500 \citep{hankrepo,hankmodel}, a source-conditional two-stage fine-tuning system for Whisper-small \citep{whisper}. Starting from a task-tuned checkpoint, external audio from WenetSpeech-Yue \citep{wsyue} updates the encoder only, competition utterances update the whole model, and optimizer steps alternate strictly one-to-one without mixing sources inside a step. Unlike a static parameter freeze, W500 toggles binary decoder masks per step inside a single AdamW optimizer, paired with a two-stage exposure schedule and checkpoint eligibility thresholds (Section~\ref{sec:method}). Alongside W500, we evaluate a converged twelve-run Whisper capacity matrix across three model sizes, full fine-tuning versus LoRA \citep{lora}, and two seeds, together with an in-matrix source-conditional ablation and controlled probes on reranking, system combination, and training duration (Section~\ref{sec:matrix} and Section~\ref{sec:sc}).

\begin{figure}[t]\centering\includegraphics[width=\linewidth]{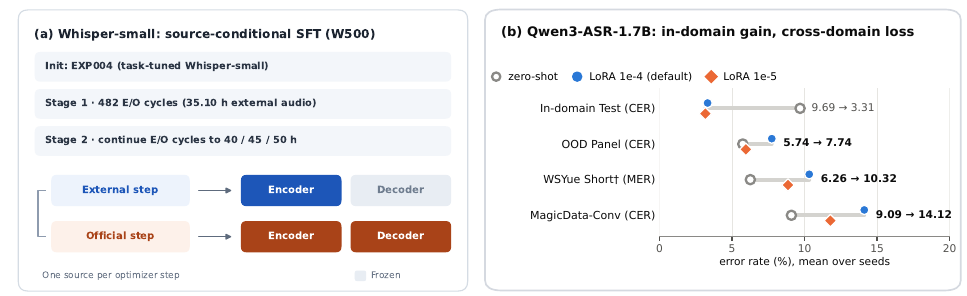}\caption{Method and empirical overview. (a) Whisper-small source-conditional fine-tuning (W500): initialized from task-tuned checkpoint EXP004, Stage 1 runs 482 external/official cycles (35.10 h cumulative external audio), and Stage 2 continues the alternation across 40, 45, and 50 h nominal targets; external steps update only the encoder, while official steps update all trainable parameters. (b) Qwen3-ASR-1.7B on four test sets: zero-shot versus the default LoRA recipe ($10^{-4}$) and the lower learning rate ($10^{-5}$), mean over three seeds (Table~\ref{tab:qwen}; WSYue Short from the cross-hardware replication, Table~\ref{tab:wsyue}). Labels give zero-shot $\rightarrow$ default recipe; bold labels mark the sets on which fine-tuning raises the error rate.}\label{fig:fig_overview_v5}\end{figure}

For a strong-prior backbone, we fine-tuned Qwen3-ASR \citep{qwen3asr}, the reference backbone of the ChinaVoices challenge \citep{chinavoices}, using that challenge's LoRA recipe on the same six hours of competition speech at 0.6B and 1.7B while varying only the learning rate. We evaluate on the competition in-domain test set, three independent two-speaker spontaneous conversation recordings from MagicData \citep{magic} held out prior to any evaluation, and, via a cross-hardware replication of all runs, the public WSYue-ASR-eval benchmark \citep{wsyue} (Section~\ref{sec:qwen}). At the default learning rate, large in-domain gains come with systematic degradation on every out-of-domain Cantonese test set across both sizes and every seed. Reducing the learning rate tenfold preserves in-domain accuracy while substantially narrowing the out-of-domain loss, and Section~\ref{sec:appb} shows that the conversational regression remains well above zero-shot even after normalizing numerals and regional Cantonese spelling variants. The Whisper and Qwen3-ASR experiments exhibit different adaptation outcomes: weak-prior models improve across all Cantonese sets, whereas strong-prior models show an in-domain versus out-of-domain trade-off of the kind reported as acoustic narrowing \citep{narrowing}. Section~\ref{sec:data} defines the datasets, scoring protocols, and the split between official competition-phase records at commit f2c576f (Section~\ref{sec:matrix}, Table~\ref{tab:matrix}, and Table~\ref{tab:sc}) and unified symmetric evaluations (Table~\ref{tab:whisper_unified} to Table~\ref{tab:foldb}), which are followed by the Whisper and Qwen3-ASR experiments in Section~\ref{sec:whisper} and Section~\ref{sec:qwen} and reproduction details in Section~\ref{sec:app}.

\section{Related Work}\label{sec:related}

Selective fine-tuning and parameter freezing. Updating only a subset of pretrained parameters is an established strategy for small target corpora, including gradual unfreezing in ULMFiT \citep{ulmfit}, surgical fine-tuning tailored to distribution-shift types \citep{surgical}, and low-rank adapters such as LoRA \citep{lora}, which yielded lower target-task accuracy than full fine-tuning on programming and mathematics while better preserving out-of-domain capabilities \citep{lorafg}. In speech recognition, Bell et al. survey neural ASR adaptation methods \citep{bell2020}, Rouditchenko et al. compare Whisper against the self-supervised XLS-R encoder on unseen languages \citep{rouditchenko}, and Field et al. investigate encoder-only and decoder-only Whisper fine-tuning \citep{field23}. Rather than applying a fixed freeze to either Whisper subnetwork, W500 toggles a binary decoder mask per optimizer step according to the data source within a single AdamW optimizer, combined with a two-stage audio-hour schedule and validation gating.

Source-conditioned optimization. Prior work conditions parameter updates on batch origin or gradient agreement via cosine-similarity gating of auxiliary gradients \citep{du18}, projection of conflicting task gradients in PCGrad \citep{pcgrad}, L2-SP regularization toward pretrained weights \citep{l2sp}, and weight-space interpolation with zero-shot weights in WiSE-FT \citep{wiseft}. W500 implements a direct source-routed hard mask without computing cross-source gradient inner products or maintaining an auxiliary weight copy during training.

Cantonese ASR and cross-domain degradation. Large Cantonese corpora such as WenetSpeech-Yue (21,800 h \citep{wsyue}) and MDCC \citep{mdcc} have supported public Cantonese adaptations, including LoRA-INT8 Whisper for edge deployment \citep{loraint8}, a Qwen3-ASR-1.7B Hong Kong Cantonese fine-tune \citep{koelony}, and the sixteen-dialect ChinaVoices LoRA baseline \citep{chinavoices}. Outside-domain losses after small-corpus fine-tuning have been documented as acoustic narrowing in Turkish and Kazakh Whisper models \citep{narrowing}, collapse--repair--drift dynamics separating early recovery from persistent Mandarin decline across learning rates in five Chinese dialects \citep{crd}, and Mandarin forgetting after dialect adaptation mitigated by on-policy self-distillation \citep{opsd}. Our Qwen3-ASR study isolates a single Cantonese competition corpus and LoRA recipe, holding training data, step count, and LoRA configuration fixed while varying only the learning rate across matched seeds, two model sizes, and two hardware setups.

\section{Data, Evaluation Sets, and Scoring}\label{sec:data}

\subsection{Datasets and Evaluation Splits}\label{sec:datasets}

The competition training set consists of 6,292 scripted Cantonese utterances (6.27 h) from Cantonese Life Scenarios with simplified-character transcripts. From the competition data, we designate a fixed validation split (Val, 702 utterances) and use the competition's 1,900-utterance public block (supplied with reference transcripts for local evaluation) as In-domain Test (stored under \texttt{public} in the repository). Official submissions are evaluated online on a separate 200-utterance hidden test set whose audio and transcripts are not public. For W500, both In-domain Test and the historical screening panel filtered anomalous Stage 2 candidate checkpoints against predefined thresholds (Section~\ref{sec:method}), so the untouched blind evaluations for W500 are the 200-utterance hidden test set and MagicData-Conv. To assess out-of-domain generalization and cross-lingual retention, we evaluate on four external or independent sets, summarized in Table~\ref{tab:datasets}:

\begin{itemize}\item Historical OOD and re-evaluation OOD panels: In Table~\ref{tab:matrix} and Table~\ref{tab:sc}, the historical screening panel (Hist. OOD) contains 1,000 utterances from Common Voice zh-HK \citep{cv} (traditional references) and 1,000 from MDCC \citep{mdcc}, both of which also contributed to the Whisper external73 training mix. In Table~\ref{tab:whisper_unified}, Table~\ref{tab:qwen}, and Figure~\ref{fig:fig_qwen_v5}, OOD Panel denotes the 2,000-utterance re-evaluation set; we did not retain an item-level mapping between this re-evaluation set and the historical screening panel, and it is completely unseen for Qwen3-ASR.\item Spontaneous conversations (MagicData-Conv) \citep{magic}: 10 two-speaker conversational recordings (20 speakers, 3,466 utterances cut at annotated boundaries) in colloquial Cantonese. Three recordings (A0006, A0018, A0032; 6 speakers; 1,110 utterances, 1.10 h) were held out prior to inspecting any conversation results and never used for training or selection.\item Mandarin retention (AISHELL-1) \citep{aishell}: a 12-speaker subset drawn from the AISHELL-1 training split, never trained on in this work, used solely to check retention of Mandarin recognition relative to each base model.\item Public Cantonese benchmark (WSYue-ASR-eval) \citep{wsyue}: Short (7,060 clips) and Long (370 recordings) splits. A text-level audit (exact string match and shared 10-grams) against our competition training and validation transcripts found zero overlap. Recording-level overlap cannot be verified from text alone.\end{itemize}

\begin{table}[t]\centering\footnotesize\setlength{\tabcolsep}{3.5pt}\caption{Summary of datasets and evaluation sets used in this report.}\label{tab:datasets}\begin{adjustbox}{max width=\linewidth,center}\begin{tabular}{lp{3.4cm}lp{6.6cm}}\toprule
dataset / evaluation set & size \& speech style & script & role and independence\\\midrule
Official Train & 6,292 utt. (6.27 h), scripted & Simplified & In-domain supervised training data\\
Val & 702 utt., scripted & Simplified & Fixed split from official data; primary ranking set for W500\\
In-domain Test & 1,900 utt., scripted & Simplified & Public competition block (\texttt{public} in repo); screened W500\\
Hidden Test & 200 utt., scripted & -- & Official platform evaluation set (transcripts not public)\\
\midrule
Hist. OOD & 2,000 utt. (1k CV + 1k MDCC) & Mixed & Screening panel in Tables 2 and 4; overlaps external73\\
OOD Panel & 2,000 utt. (re-eval panel) & Mixed & Re-evaluation panel in Tables 3 and 5; unseen for Qwen3-ASR\\
MagicData-Conv & 3 held-out of 10 recordings (1,110 utt., 1.10 h), conv. & Colloquial Yue & Independent spontaneous speech; locked prior to evaluation\\
AISHELL-1 (Mandarin) & 12 speakers (subset), read & Simplified & Untrained Mandarin retention check relative to base model\\
WSYue-ASR-eval & Short: 7,060 clips; Long: 370 rec. & Mixed (MER) & Public benchmark; zero text/10-gram overlap with train/val\\
\bottomrule\end{tabular}\end{adjustbox}\end{table}

\subsection{Scoring Protocols and Orthographic Variants}\label{sec:scoring}

Official scoring. Following the competition evaluator, punctuation and spaces are stripped, only the hypothesis is converted to simplified characters with OpenCC \citep{opencc}, and the reference is left unchanged before computing character error rate (CER) and tol2 (the fraction of utterances with at most two character edits). This protocol is exact on Val and In-domain Test, whose references are simplified, but penalizes traditional-character differences on Hist. OOD.

Symmetric scoring. For all newly computed CER figures in this report (unless marked official), both hypothesis and reference are converted to simplified characters with OpenCC and stripped of punctuation and spaces. On the AISHELL-1 subset, Arabic and Chinese numerals are also unified with cn2an \citep{cn2an}.

WSYue-ASR-eval MER. Because the official WSYue-ASR-eval scoring script is unpublished, we re-implemented the benchmark's mixed error rate (MER): punctuation stripped, OpenCC traditional-to-simplified conversion, cn2an digit-to-Chinese-numeral conversion, English lower-casing, and one token per CJK character or English word. As a calibration check, our implementation scores SenseVoice-small-Yue \citep{sensevoice} at 5.41\% / 8.84\% MER (Short / Long), exceeding the published 5.23\% / 8.63\% \citep{wsyue} by 0.18 and 0.21 percentage points.

Cantonese orthographic variants. Written conventions for colloquial Cantonese expressions vary across corpora (e.g., \raisebox{-0.28em}{\includegraphics[height=1.15em]{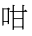}} / \raisebox{-0.28em}{\includegraphics[height=1.15em]{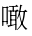}}, \raisebox{-0.28em}{\includegraphics[height=1.15em]{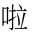}} / \raisebox{-0.28em}{\includegraphics[height=1.15em]{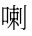}}, and \raisebox{-0.28em}{\includegraphics[height=1.15em]{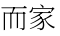}} / \raisebox{-0.28em}{\includegraphics[height=1.15em]{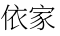}}). Our main tables preserve these distinctions without variant folding, so cross-corpus CER combines acoustic misrecognition with regional spelling differences. Section~\ref{sec:appb} quantifies sensitivity to an additional protocol that normalizes numerals and folds 15 written variants.

\subsection{Result Provenance}\label{sec:prov}

We separate two measurement sources throughout the paper. Official-scored results from the competition phase (the hidden-test platform score of W500 in Section~\ref{sec:matrix}, Table~\ref{tab:matrix}, and Table~\ref{tab:sc}) are taken directly from our repository at commit f2c576f. All unified Whisper re-evaluations (Table~\ref{tab:whisper_unified}), all Qwen3-ASR evaluations (Table~\ref{tab:qwen} and Table~\ref{tab:wsyue}), and the normalization sensitivity analysis (Table~\ref{tab:foldb}) use symmetric scoring and can be reproduced on a CPU from the released per-utterance prediction files (Section~\ref{sec:app}).

\section{Weak-Prior Adaptation: Whisper W500 and Capacity Study}\label{sec:whisper}

\subsection{Source-Conditional Two-Stage Fine-Tuning (W500)}\label{sec:method}

Let $\theta = (\theta_{enc}, \theta_{dec})$ denote the trainable Whisper parameters, where $\theta_{dec}$ includes the output projection \texttt{proj\_\allowbreak{}out} tied to the decoder embedding. Training draws optimizer steps from two sources: an external pool $E$ (500 h of quality-filtered WenetSpeech-Yue audio) and the official competition set $O$ (6,292 utterances). Every optimizer step $t$, including both gradient-accumulation microbatches, uses a single source $s_t \in \{E, O\}$ and applies a binary parameter-group mask:

\begin{equation}
m_t(\theta_{enc}) = 1\ (\forall s_t),\qquad m_t(\theta_{dec}) = \begin{cases} 1, & s_t = O\\ 0, & s_t = E \end{cases}
\label{eq:mask}
\end{equation}

yielding the active gradient components $g_t = m_t \odot \nabla_\theta L(\theta; B_t)$ under standard transcript cross-entropy loss $L$, with each microbatch contributing half of its cross-entropy loss to the accumulated gradient. Figure~\ref{fig:fig_overview_v5}(a) and Algorithm~\ref{alg:w500} summarize the procedure.

Mask behavior under AdamW. On external steps ($s_t = E$), we set \texttt{requires\_\allowbreak{}grad} of $\theta_{dec}$ to False and, as an additional guard, set the decoder parameter groups' learning rate to zero. Before gradient accumulation, gradients are cleared with \texttt{optimizer.zero\_\allowbreak{}grad(set\_\allowbreak{}to\_\allowbreak{}none=True)}, leaving frozen decoder parameters with \texttt{grad=None}. PyTorch AdamW skips these parameters entirely: their weights, first and second moments, and per-parameter step counters remain unchanged, and no decoupled weight decay is applied. Decoder updates resume on the next official step, while one shared AdamW optimizer per training run carries the encoder's optimizer state across both sources (re-initialized only when starting a new continuation branch). Global gradient-norm clipping (threshold 1.0) operates only over parameters with gradients, so external steps clip encoder gradients alone.

Optimizer and hyperparameters. Parameters are partitioned into four groups ({enc, dec} $\times$ {decay, no-decay}, excluding biases and layer-normalization parameters from decay) under AdamW with $\beta = (0.9, 0.999)$, $\epsilon = 10^{-8}$, and weight decay 0.01. The released configuration uses source-specific constant learning rates $\eta_E = 10^{-6}$ on external steps and $\eta_O = 5 \times 10^{-7}$ on official steps, effective batch size 16 (two microbatches of 8), bf16 precision, and no SpecAugment.

Two-stage schedule. For the released model, Stage 1 starts from a Whisper-small checkpoint previously fine-tuned on competition data (EXP004, \texttt{FULL\_\allowbreak{}LR1E6/checkpoint-\allowbreak{}150}). From this checkpoint, each cycle pairs one encoder-only external step ($E$) with one full-model official step ($O$), in that order. Stage 1 comprises 482 $E$ steps and 482 $O$ steps, reaching 35.10 h of cumulative WenetSpeech-Yue exposure. Stage 2 continues this 1 : 1 alternation across nominal cumulative external-audio targets of 40, 45, and 50 h (measured from the Stage 1 starting point), with intermediate save targets every 1.25 h within each 5 h window. The trainer precomputes the complete external-batch boundary nearest each target and saves checkpoints after the $O$ step of that cycle. After the 40 h and 45 h rounds, the two highest-ranked eligible checkpoints are selected as parents for the next round. The released checkpoint is \texttt{checkpoint-\allowbreak{}wenet-\allowbreak{}50h}. This schedule broadens acoustic coverage with external audio while using short competition utterances to refine character choice and control insertions, repetitions, and end-of-sequence behavior.

\begin{algorithm}[t]\caption{W500: source-conditional two-stage fine-tuning of Whisper-small.}\label{alg:w500}
\begin{lstlisting}[basicstyle=\ttfamily\footnotesize,frame=none,mathescape=true]
Input:  task-tuned Whisper-small EXP004 (FULL_LR1E6/checkpoint-150); external stream D_E (500 h
        WenetSpeech-Yue pool); official stream D_O (6,292 utt.); constant rates $\eta_E = 10^{-6}$, $\eta_O = 5 \times 10^{-7}$;
        Stage-1 budget 482 cycles; Stage-2 targets 40, 45, 50 h.
Output: selected parent checkpoints at 40 and 45 h and the release checkpoint-wenet-50h.

# One optimizer step(s):
  clear gradients with optimizer.zero_grad(set_to_none=True)
  if s == E: requires_grad(enc)=True, requires_grad(dec)=False; lr(enc)=$\eta_E$, lr(dec)=0
  else:      requires_grad(all)=True; lr(all)=$\eta_O$
  accumulate 0.5 * CE loss over two 8-utterance microbatches from source s
  clip global norm (parameters with gradients) to 1.0; AdamW.step()

# Stage 1 (init from EXP004):
  initialise one AdamW optimizer (4 groups, betas=(0.9, 0.999), eps=1e-8, wd=0.01)
  for 482 cycles: step(E); step(O)                       # 35.10 h cumulative external audio

# Stage 2 (continuation windows targeting 40, 45, 50 h):
  for each window target in {40, 45, 50} h:
      re-initialise AdamW from the selected parent(s); set save targets every 1.25 h
      while cumulative external hours < window target:
          step(E); step(O)
          if the external cursor hits the nearest complete-batch target: save checkpoint after step(O)
      screen candidates on Val, In-domain Test, Hist. OOD; rank eligible runs on Val
      select the top 2 parents at 40 h and 45 h; release checkpoint-wenet-50h at 50 h
\end{lstlisting}\end{algorithm}

Selection gates and decoding configuration. Stage 2 continuation candidates are screened against predefined thresholds on Val, In-domain Test, and Hist. OOD, covering tol2, public-set insertion counts, severe errors, runaway repetition, Unicode replacement characters (U+FFFD), maximum-length hits, and OOD CER. Eligible candidates are ranked on Val by the 90-point accuracy proxy $70(1 - \mathrm{CER}) + 20\,\mathrm{tol2}$ (with both metrics as fractions, omitting the competition's constant 10-point offset, which does not affect ordering). Score differences below 0.05 break ties by lower Val CER, higher Val tol2, fewer severe Val errors, shorter cumulative external exposure, and lexicographic label. Because In-domain Test and Hist. OOD thresholds could reject candidates, both sets contributed to model selection. The standalone submission script uses \texttt{language="zh"}, \texttt{task="transcribe"}, single-beam greedy decoding, \texttt{max\_\allowbreak{}length=225}, \texttt{no\_\allowbreak{}repeat\_\allowbreak{}ngram\_\allowbreak{}size=4}, \texttt{repetition\_\allowbreak{}penalty=1.05}, and the Whisper suppression lists.

\subsection{W500 Submission Results, Capacity Matrix, and Controlled Probes}\label{sec:matrix}

W500 submission results. On the platform's 200-utterance hidden test set (evaluated online with the single-beam greedy submission script and official platform scoring), W500 achieved an official score of 69.49 (24.73\% CER and 0.340 tol2) \citep{hankrepo}, matching the task's published 100-point scoring formula $((1 - \mathrm{CER}) \times 70\% + \mathrm{tol2} \times 20\% + 10\%) \times 100 = 70(1 - 0.2473) + 20(0.340) + 10 = 69.49$ \citep{dimsum}, with no final leaderboard snapshot saved. Under our unified 2-beam symmetric protocol, W500 achieved 8.09\% CER on the 1,900-utterance In-domain Test block (Table~\ref{tab:whisper_unified}, which also filtered Stage 2 candidates) and lowered MagicData-Conv CER by 37.65 percentage points relative to zero-shot Whisper-small (from 62.69\% to 25.04\%).

Converged capacity matrix. Our main Whisper controlled comparison evaluates full fine-tuning and LoRA across three model sizes (small, medium, and large-v2, two seeds each: 42 and 43) for 3 epochs (step 4,371) on the external73 mix. That mix is a training recipe with 73\% external data: 23,304 shuffled utterances combining Official Train, Common Voice zh-HK train, and MDCC. Training uses a cosine learning-rate schedule and pins transformers 4.57.6, because 5.x versions alter scores. Table~\ref{tab:matrix} reports the official competition-report results at commit f2c576f \citep{hankrepo}, and Table~\ref{tab:whisper_unified} re-evaluates all twelve checkpoints alongside zero-shot baselines and the released W500 model under unified symmetric scoring and 2-beam decoding. Tables 2 and 3 show four consistent empirical patterns:

\begin{table}[t]\centering\footnotesize\setlength{\tabcolsep}{3.5pt}\caption{Capacity matrix results at step 4,371 from the competition report at commit f2c576f \citep{hankrepo} (Section~\ref{sec:app}). Val (702 utt.) and In-domain Test (1,900 utt., labeled \texttt{public} in the repo) use official scoring; Hist. OOD (1,000 Common Voice zh-HK + 1,000 MDCC) shows normalized CER with both sides simplified (unnormalized CER was 32--35\%). Dashes (--) mark entries omitted from that summary table; all six arms are evaluated in full in Table~\ref{tab:whisper_unified}. W500 reference on Val: tol2 0.855, CER 8.59\%.}\label{tab:matrix}\begin{adjustbox}{max width=\linewidth,center}\begin{tabular}{lrrrr}\toprule
arm & Val tol2 (s42 / s43) & Val CER \% (s42 / s43) & In-domain Test CER \% (s42 / s43) & Hist. OOD norm. CER \% (s42 / s43)\\\midrule
large-v2, full & 0.895 / 0.895 & 6.85 / 6.77 & 5.74 / 5.58 & 6.31 / 6.50\\
medium, full & 0.892 / 0.887 & 7.36 / 6.97 & 6.26 / 7.39 & 7.12 / 7.13\\
large-v2, LoRA & 0.877 / 0.872 & 7.99 / 7.96 & 5.07 / 5.10 & 6.38 / 6.66\\
small, full & 0.826 / 0.829 & 9.35 / 9.47 & 9.13 / 9.53 & 9.47 / 9.47\\
medium, LoRA & 0.822 / 0.819 & 9.43 / 9.69 & -- & --\\
small, LoRA & 0.701 / 0.729 & 14.02 / 13.63 & -- & --\\
\bottomrule\end{tabular}\end{adjustbox}\end{table}

\begin{table}[t]\centering\footnotesize\setlength{\tabcolsep}{3.5pt}\caption{Unified symmetric re-evaluation of Whisper models: symmetric CER (\%, both sides simplified, punctuation removed) under 2-beam decoding, averaged over two seeds for matrix checkpoints. OOD Panel denotes the 2,000-utterance re-evaluation set (Section~\ref{sec:data}); AISHELL-1 uses cn2an numeral normalization.}\label{tab:whisper_unified}\begin{adjustbox}{max width=\linewidth,center}\begin{tabular}{lrrrrr}\toprule
checkpoint & Val & In-domain Test & OOD Panel & MagicData-Conv & AISHELL-1 (Mandarin)\\\midrule
Whisper-small, zero-shot & 39.26 & 40.02 & 34.20 & 62.69 & 9.84\\
  full SFT, step 4371 (mean of 2 seeds) & 9.34 & 9.34 & 9.46 & 28.46 & 12.40\\
  LoRA, step 4371 (mean of 2 seeds) & 13.80 & 12.76 & 12.65 & 29.61 & 12.45\\
\midrule
Whisper-medium, zero-shot & 35.46 & 33.16 & 29.26 & 59.83 & 6.87\\
  full SFT, step 4371 (mean of 2 seeds) & 7.11 & 6.80 & 7.12 & 25.13 & 8.99\\
  LoRA, step 4371 (mean of 2 seeds) & 9.53 & 6.41 & 9.98 & 26.05 & 7.02\\
\midrule
Whisper-large-v2, zero-shot & 31.78 & 34.26 & 27.73 & 58.75 & 5.56\\
  full SFT, step 4371 (mean of 2 seeds) & 6.74 & 5.61 & 6.43 & 23.14 & 7.64\\
  LoRA, step 4371 (mean of 2 seeds) & 7.90 & 5.07 & 6.51 & 23.38 & 5.26\\
\midrule
Released W500 model (Whisper-small) & 8.51 & 8.09 & 8.85 & 25.04 & 11.77\\
\bottomrule\end{tabular}\end{adjustbox}\end{table}

\begin{itemize}\item Validation gains diminish beyond medium: Averaging Val tol2 across two seeds for full fine-tuning in Table~\ref{tab:matrix} yields gains of +0.062 from small to medium and only +0.005 from medium to large-v2 despite roughly doubling parameter count. In Table~\ref{tab:whisper_unified}, large-v2 full improves over medium full by 0.37 CER on Val and 1.19 CER on In-domain Test.\item Validation and in-domain test rankings diverge: Across all six configurations, the Spearman rank correlation between Val and In-domain Test is 0.60 both for two-seed mean tol2 in the competition report \citep{hankrepo} and for two-seed mean symmetric CER in Table~\ref{tab:whisper_unified}. LoRA trails full fine-tuning on Val at every scale, yet outperforms full fine-tuning on In-domain Test at both medium (6.41 vs. 6.80) and large-v2 (5.07 vs. 5.61). Only the medium and large-v2 LoRA arms move up in rank. Model selection followed our validation-based rule, which ranks medium and large-v2 LoRA below their full fine-tuning counterparts.\item Unnormalized OOD errors reflect script mismatch: Converting both references and predictions to simplified characters reduces Hist. OOD CER by 72.6\%--80.6\% across the four evaluated arms relative to unnormalized scoring. For large-v2 full, normalized Hist. OOD CER sits within 0.6--0.9 points of In-domain Test (6.31\% / 6.50\% vs. 5.74\% / 5.58\% in Table~\ref{tab:matrix}). Table~\ref{tab:whisper_unified} gives a similar 6.43\% mean (6.33\% and 6.52\% per seed).\item Residual errors form a long tail of character substitutions: At step 4,371 (seed 42), large-v2 full makes 641 edits across 702 Val utterances, including 454 character substitutions (70.8\%). The 50 worst utterances account for 243 edits (37.9\%, leaving 398 edits or 62.1\% across the rest), and two utterances meet the $\geq$ 80\% edit-distance screening threshold for possible audio--transcript misalignment. On the 268 utterances where both large-v2 full and small full make errors, those shared utterances account for 575 of large-v2's 641 edits (89.7\%).\end{itemize}

Several additional experiments failed toimprove recognition accuracy. Full fine-tuning of Whisper-small on external73 trailed W500 on both Val and In-domain Test. A character 5-gram reranker selected weight zero on validation, MBR and ROVER static fusion trailed the best single checkpoint, the 5-best oracle ceiling added only +0.003 tol2, and extending medium LoRA from 3 to 5 epochs had little effect on Val CER. Two factors warrant further investigation: upgrading from transformers 4.57.6 to 5.6.2 raises two-seed mean Val tol2 for small full by roughly +0.02 (excluded to keep the library version fixed across the matrix), and 29\%--35\% of Cantonese reference character occurrences split into multiple tokens under Whisper's default tokenizer.

\subsection{In-Matrix Source-Conditional Ablation}\label{sec:sc}

Within the Section~\ref{sec:matrix} capacity matrix, we test a source-conditional (SC) variant of full fine-tuning at small and large-v2 using the same external73 mix, cosine schedule, and single-pass epoch exposure (6,292 official and 17,012 external utterances, yielding $\approx$2.70 external steps per official step rather than W500's 1 : 1 alternation). Whereas full optimizer steps contain 16 utterances, SC uses microbatches of 4 to accommodate the four-utterance tail of each stream. For small, we train a microbatch-4 mixed-data baseline for a microbatch-matched comparison, and at large-v2, both SC and the baseline configuration specify microbatches of 4 (Table~\ref{tab:sc}).

\begin{table}[t]\centering\footnotesize\setlength{\tabcolsep}{3.5pt}\caption{In-matrix source-conditional (SC) ablation versus mixed-data full fine-tuning from the competition report at commit f2c576f \citep{hankrepo} (\url{https://github.com/Vanxun-Hank/cantonese-asr/blob/f2c576f7da5d376219bea4246641730713a762a1/reports/raw_winner_p2_full_converged.md}). small compares validation-selected checkpoints (the seed-43 microbatch-4 baseline diverged at the last checkpoint, so step 3,643 is selected). Both large-v2 configurations specify microbatches of 4 and are evaluated at step 4,371; the last row shows exact per-seed tol2 differences under original Hist. OOD scoring.}\label{tab:sc}\begin{adjustbox}{max width=\linewidth,center}\begin{tabular}{lrr}\toprule
comparison & seed 42 & seed 43\\\midrule
small, Val: SC - microbatch-4 baseline (tol2) & -0.017 & -0.023\\
small, Val: SC - microbatch-4 baseline (CER points) & +0.91 & +0.43\\
small, Val: microbatch-4 - microbatch-8 baseline (tol2) & +0.001 & -0.007\\
\midrule
large-v2, In-domain Test: tol2 (SC / baseline) & 0.950 / 0.947 & 0.951 / 0.945\\
large-v2, In-domain Test: CER \% (SC / baseline) & 5.20 / 5.74 & 5.25 / 5.58\\
large-v2, original Hist. OOD: SC - baseline (tol2) & -0.0075 & -0.0040\\
\bottomrule\end{tabular}\end{adjustbox}\end{table}

At small capacity, SC yields lower Val tol2 and higher Val CER (+0.91 and +0.43 points) than the microbatch-matched baseline in both seeds, while the microbatch-4 versus microbatch-8 baseline difference is small and changes sign, and the In-domain Test comparison at small is inconclusive. At large-v2, SC trails slightly on Val (-0.006 and -0.001 tol2) yet achieves higher In-domain Test tol2 and 9.5\% / 6.0\% relative CER reductions (5.20\% vs. 5.74\% and 5.25\% vs. 5.58\%). This is the third case, alongside LoRA and model scale, where Val and In-domain Test rank arms differently. Under original Hist. OOD scoring, SC scores slightly below the baseline at both sizes and seeds (-0.0075 and -0.0040 at large-v2), reflecting that protocol's sensitivity to script mismatch. Because Table~\ref{tab:sc} changes source homogeneity and external-step decoder freezing together on external73, it isolates neither factor alone nor W500's WenetSpeech-Yue pool and two-stage schedule.

\subsection{Matched Controls on MDCC External Data}\label{sec:ctrl}

The ablation in Table~\ref{tab:sc} cannot separate single-source stepping from decoder freezing, and its two seeds leave the size of the small-model effect uncertain. We therefore ran a matched control in our own trainer, with MDCC as the external source instead of the external73 mix or W500's WenetSpeech-Yue pool. Whisper-small is initialized from one shared encoder-only checkpoint (20 h of MDCC audio), then trained with alternating single-source steps to 35 h of external exposure on the same 6,292 official utterances, with the same batch size and external exposure in every arm. Three update rules are compared: the W500 rule (decoder frozen on external steps), the same single-source alternation with the decoder trainable on external steps (alt-full), and conventional mixed-batch fine-tuning in which every microbatch is half external and half official. The W500 rule and alt-full share one schedule (constant $10^{-6}$ on external steps and $5 \times 10^{-7}$ on official steps, 4,749 optimizer steps) and differ only in the decoder mask. The mixed arm uses the official rate $5 \times 10^{-7}$ on every step (4,750 steps), so comparisons with it are between two training recipes and do not isolate batch composition from learning rate. Each rule was run with three seeds and evaluated on all sets (Table~\ref{tab:roundy}). We tested two hypotheses: that the decoder freeze costs Val accuracy, and that the W500 rule does not reduce MagicData-Conv CER relative to conventional fine-tuning at equal exposure.

\begin{table}[t]\centering\footnotesize\setlength{\tabcolsep}{3.5pt}\caption{Matched controls in our trainer (Whisper-small, MDCC external audio, shared initialization, 35 h external exposure; W500 rule and alt-full share all hyperparameters, the mixed arm uses $5 \times 10^{-7}$ on every step): symmetric CER (\%) and WSYue MER (\%), mean $\pm$ sample standard deviation over seeds 42, 43, 44 (per-seed values in the project log). This setting differs from Table~\ref{tab:sc} (external73, cosine schedule) and from the released W500 run (WenetSpeech-Yue pool).}\label{tab:roundy}\begin{adjustbox}{max width=\linewidth,center}\begin{tabular}{lrrrrrr}\toprule
update rule (3 seeds) & Val & In-dom. Test & OOD Panel & MagicData-Conv & WSYue Short & WSYue Long\\\midrule
W500 rule (decoder frozen on ext. steps) & 11.29 $\pm$ 0.15 & 9.86 $\pm$ 0.06 & 15.68 $\pm$ 0.24 & 29.36 $\pm$ 0.10 & 20.45 $\pm$ 0.08 & 17.19 $\pm$ 0.26\\
alt-full (single-source, decoder trainable) & 10.69 $\pm$ 0.22 & 9.43 $\pm$ 0.12 & 14.20 $\pm$ 0.07 & 28.68 $\pm$ 0.35 & 19.17 $\pm$ 0.03 & 15.21 $\pm$ 0.13\\
mixed-batch fine-tuning & 11.17 $\pm$ 0.12 & 9.76 $\pm$ 0.09 & 14.53 $\pm$ 0.05 & 28.71 $\pm$ 0.12 & 19.45 $\pm$ 0.09 & 15.46 $\pm$ 0.14\\
\bottomrule\end{tabular}\end{adjustbox}\end{table}

The results support both hypotheses. Freezing the decoder costs accuracy: alt-full has lower Val CER than the W500 rule in every seed (10.69 vs. 11.29), and lower MER on both WSYue splits in every seed (19.17 vs. 20.45 Short; 15.21 vs. 17.19 Long). It also has lower MagicData-Conv CER in every seed (28.68 vs. 29.36), although the conversation difference (0.67 points) is within twice the larger seed spread (0.35). Relative to the mixed-batch recipe at the same data and external exposure, the W500 rule does not lower MagicData-Conv CER (29.36 vs. 28.71, higher in every seed) or In-domain Test CER (9.86 vs. 9.76). Alt-full, which keeps single-source steps but updates the decoder on both sources, has the lowest Val CER and WSYue MER of the three rules and matches the mixed recipe on MagicData-Conv. Attributing these differences to batch composition alone would require a mixed arm at the alternating arms' learning rates, which we did not run. These controls use MDCC external audio and constant learning rates, so they speak to the decoder mask inside a W500-style schedule and not to the released run's WenetSpeech-Yue pool; they do not reproduce Table~\ref{tab:sc}, and the released model's gains over its zero-shot base (Table~\ref{tab:whisper_unified}) reflect the full adaptation path rather than the mask.

\section{Cross-Domain Trade-Offs on Qwen3-ASR}\label{sec:qwen}

\subsection{Setup and Main Results}\label{sec:setup}

\begin{figure}[t]\centering\includegraphics[width=\linewidth]{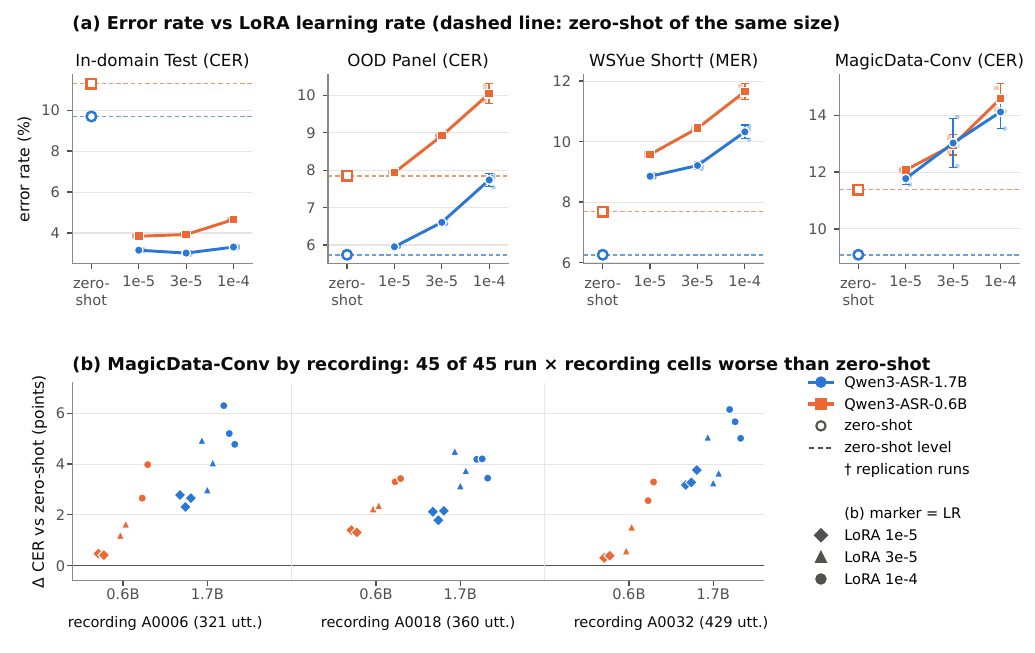}\caption{Qwen3-ASR fine-tuning on six hours of scripted speech. (a) Error rate versus LoRA learning rate for 0.6B and 1.7B: symmetric CER on In-domain Test, the 2,000-utterance OOD Panel, and MagicData-Conv from the original runs (Table~\ref{tab:qwen}), and MER on WSYue-ASR-eval Short from the cross-hardware replication (Table~\ref{tab:wsyue}). Hollow markers and dashed lines give the zero-shot model of the same size; error bars show sample standard deviation and small dots individual seeds. (b) Change in CER relative to zero-shot on each of the three held-out MagicData-Conv recordings for all 15 fine-tuned runs; marker shape gives the learning rate. Every point lies above zero.}\label{fig:fig_qwen_v5}\end{figure}

To test whether six hours of scripted competition fine-tuning degrade spontaneous speech when the pretrained backbone is already strong in Cantonese, we fine-tuned Qwen3-ASR-0.6B (two seeds: 42, 43) and Qwen3-ASR-1.7B (three seeds: 42, 43, 44) on the same 6,292 competition utterances. We used the ChinaVoices LoRA recipe: LoRA rank 8, $\alpha = 32$, dropout 0.05 applied only to the text decoder with the speech encoder frozen. Training uses AdamW with $\beta = (0.9, 0.95)$, weight decay 0.1, cosine schedule with 5\% warm-up, global batch 16, 3 epochs (1,182 steps), and bf16 precision. Holding training data, step count, LoRA configuration, and optimizer settings fixed, we vary only the learning rate: $10^{-4}$ (the challenge recipe), $3 \times 10^{-5}$, and $10^{-5}$. Table~\ref{tab:qwen} and Figure~\ref{fig:fig_qwen_v5} report symmetric CER across all 17 original runs (2 zero-shot and 15 fine-tuned), matching training-side logs to the second decimal. Both model sizes exhibit the followingpatterns:

\begin{table}[t]\centering\footnotesize\setlength{\tabcolsep}{3.5pt}\caption{Qwen3-ASR fine-tuned on six hours of competition speech: symmetric CER (\%). Fine-tuned arms report mean $\pm$ sample standard deviation over 2 seeds (0.6B) or 3 seeds (1.7B); zero-shot is a single evaluation.}\label{tab:qwen}\begin{adjustbox}{max width=\linewidth,center}\begin{tabular}{lrrrr}\toprule
model & recipe & In-domain Test & MagicData-Conv & OOD Panel\\\midrule
Qwen3-ASR-0.6B & zero-shot & 11.30 & 11.38 & 7.85\\
Qwen3-ASR-0.6B & LoRA, lr $10^{-4}$ (challenge recipe) & 4.66 $\pm$ 0.03 & 14.58 $\pm$ 0.53 & 10.05 $\pm$ 0.27\\
Qwen3-ASR-0.6B & LoRA, lr $3 \times 10^{-5}$ & 3.93 $\pm$ 0.02 & 12.96 $\pm$ 0.36 & 8.92 $\pm$ 0.03\\
Qwen3-ASR-0.6B & LoRA, lr $10^{-5}$ & 3.84 $\pm$ 0.02 & 12.08 $\pm$ 0.02 & 7.94 $\pm$ 0.02\\
\midrule
Qwen3-ASR-1.7B & zero-shot & 9.69 & 9.09 & 5.74\\
Qwen3-ASR-1.7B & LoRA, lr $10^{-4}$ (challenge recipe) & 3.31 $\pm$ 0.06 & 14.12 $\pm$ 0.58 & 7.74 $\pm$ 0.17\\
Qwen3-ASR-1.7B & LoRA, lr $3 \times 10^{-5}$ & 3.02 $\pm$ 0.07 & 13.02 $\pm$ 0.86 & 6.61 $\pm$ 0.07\\
Qwen3-ASR-1.7B & LoRA, lr $10^{-5}$ & 3.16 $\pm$ 0.05 & 11.77 $\pm$ 0.20 & 5.95 $\pm$ 0.00\\
\bottomrule\end{tabular}\end{adjustbox}\end{table}

\begin{enumerate}\item Large in-domain gains come with systematic out-of-domain loss. At the challenge learning rate ($10^{-4}$), 1.7B lowers In-domain Test CER from 9.69\% to 3.31\% while raising MagicData-Conv CER from 9.09\% to 14.12\%, with every seed worse than zero-shot. The loss is not driven by one recording: each of the 15 fine-tuned runs, at every learning rate and both sizes, has higher CER than its zero-shot model on each of the three MagicData-Conv recordings (45 of 45 run--recording pairs, Figure~\ref{fig:fig_qwen_v5}b). Because the set contains only three independent recordings, this shows consistency across runs rather than across many speakers or sessions. At 0.6B, In-domain Test improves from 11.30\% to 4.66\% while MagicData-Conv degrades from 11.38\% to 14.58\% in both seeds. OOD Panel moves in the same direction at both sizes.\item Lowering the learning rate monotonically reduces the loss without hurting in-domain accuracy. On both out-of-domain sets, CER decreases monotonically from $10^{-4}$ to $3 \times 10^{-5}$ to $10^{-5}$ at both model sizes. On MagicData-Conv, the degradation over zero-shot shrinks from 5.03 to 2.68 CER points at 1.7B ($\approx$47\% reduction) and from 3.20 to 0.69 points based on unrounded seed means at 0.6B (0.70 from rounded table entries, roughly a 78\% reduction). Meanwhile, in-domain CER at 1.7B stays within 0.3 points across all three learning rates (3.31, 3.02, and 3.16, with seed spread 0.05--0.07), and at 0.6B lower rates perform better (4.66 down to 3.84). Lowering the learning rate does not eliminate the out-of-domain loss, as every fine-tuned arm remains above zero-shot on MagicData-Conv, but it shows that $10^{-4}$ is higher than needed for six hours of scripted speech.\end{enumerate}

\subsection{Cross-Hardware Replication and Public Benchmark WSYue-ASR-eval}\label{sec:rep}

Cross-hardware replication. Because the original 17 Qwen3-ASR runs did not save model weights, we retrained all 17 configurations with identical data, recipe, step counts, and seeds on different GPU hardware (torch 2.10, transformers 5.24.1) and re-evaluated all sets alongside WSYue-ASR-eval (Table~\ref{tab:wsyue}). Comparing the 17 matched individual runs (15 fine-tuned seeds and 2 zero-shot evaluations) before seed averaging, the maximum absolute discrepancy relative to the original runs is 0.15 CER on In-domain Test, 0.24 on OOD Panel (recomputed from \texttt{pred\_\allowbreak{}ood.jsonl}, not tabulated in Table~\ref{tab:wsyue}), and 0.67 on MagicData-Conv. For 16 of the 17 individual runs, the difference is within 0.5 CER on all three sets. On the seed-averaged entries reported in Table~\ref{tab:wsyue}, the largest mean difference is 0.09 CER on In-domain Test and 0.29 on MagicData-Conv, preserving the observed learning-rateordering and cross-domain degradation.

\begin{table}[t]\centering\footnotesize\setlength{\tabcolsep}{3.5pt}\caption{Cross-hardware replication (torch 2.10; identical data, recipe, steps, and seeds) and public WSYue-ASR-eval benchmark. In-domain Test and MagicData-Conv show symmetric CER \% (orig. / repl.); WSYue columns show MER \% of the replicated runs ($\pm$ sample standard deviation over 2 seeds at 0.6B or 3 seeds at 1.7B; zero-shot is a single run).}\label{tab:wsyue}\begin{adjustbox}{max width=\linewidth,center}\begin{tabular}{lrrrrr}\toprule
model & recipe & In-domain Test (orig. / repl.) & MagicData-Conv (orig. / repl.) & WSYue Short MER & WSYue Long MER\\\midrule
Qwen3-ASR-0.6B & zero-shot & 11.30 / 11.21 & 11.38 / 11.39 & 7.68 & 10.45\\
Qwen3-ASR-0.6B & LoRA, lr $10^{-4}$ & 4.66 / 4.73 & 14.58 / 14.65 & 11.65 $\pm$ 0.27 & 13.20 $\pm$ 0.59\\
Qwen3-ASR-0.6B & LoRA, lr $3 \times 10^{-5}$ & 3.93 / 3.96 & 12.96 / 13.00 & 10.44 $\pm$ 0.04 & 11.68 $\pm$ 0.05\\
Qwen3-ASR-0.6B & LoRA, lr $10^{-5}$ & 3.84 / 3.89 & 12.08 / 12.06 & 9.58 $\pm$ 0.09 & 11.02 $\pm$ 0.01\\
\midrule
Qwen3-ASR-1.7B & zero-shot & 9.69 / 9.69 & 9.09 / 9.14 & 6.26 & 9.72\\
Qwen3-ASR-1.7B & LoRA, lr $10^{-4}$ & 3.31 / 3.23 & 14.12 / 14.11 & 10.32 $\pm$ 0.23 & 13.71 $\pm$ 0.19\\
Qwen3-ASR-1.7B & LoRA, lr $3 \times 10^{-5}$ & 3.02 / 3.05 & 13.02 / 13.31 & 9.21 $\pm$ 0.12 & 11.24 $\pm$ 0.33\\
Qwen3-ASR-1.7B & LoRA, lr $10^{-5}$ & 3.16 / 3.15 & 11.77 / 11.86 & 8.85 $\pm$ 0.06 & 10.37 $\pm$ 0.03\\
\bottomrule\end{tabular}\end{adjustbox}\end{table}

Public benchmark results. On WSYue-ASR-eval (which shares zero exact matches or 10-grams with the competition training or validation text), every fine-tuned arm across both sizes and all seeds is worse than zero-shot on both Short and Long splits, and the loss shrinks monotonically with the learning rate. At 1.7B, Short MER moves from 6.26 zero-shot to 10.32 ($10^{-4}$), 9.21 ($3 \times 10^{-5}$), and 8.85 ($10^{-5}$), and Long MER from 9.72 to 13.71, 11.24, and 10.37. At 0.6B, Short moves from 7.68 to 11.65, 10.44, and 9.58, and Long from 10.45 to 13.20, 11.68, and 11.02. Choosing $10^{-5}$ removes 36\% of the Short MER increase at 1.7B and 52\% at 0.6B.

\subsection{Scope and Comparison with Whisper}\label{sec:scope}

Our Whisper models do not exhibit this inverse trade-off in Cantonese because their zero-shot Cantonese CER starts high (ranging from 27.73\% to 62.69\% across the three model sizes and four Cantonese sets). Fine-tuning on the competition mix improves Whisper on every Cantonese set (e.g., 62.69\% to 28.46\% on MagicData-Conv for small full), even though conversational error remains much higher than scripted error (23\%--30\% vs. 5\%--14\%). Among the Whisper-small entries in Table~\ref{tab:whisper_unified}, the released W500 model also has lower CER than the two three-epoch mixed-data fine-tunes on both evaluation sets (8.09 vs. 9.34 full and 12.76 LoRA on In-domain Test; 25.04 vs. 28.46 and 29.61 on MagicData-Conv), although the three systems differ in training data, schedule, and exposure, so this is an observation rather than a matched comparison. By contrast, on AISHELL-1 (Mandarin), where Whisper's zero-shot CER is already low (5.56\%--9.84\% in Table~\ref{tab:whisper_unified}), full fine-tuning degrades Mandarin retention by 2.08--2.56 percentage points across all three sizes (9.84\% to 12.40\% at small, 6.87\% to 8.99\% at medium, and 5.56\% to 7.64\% at large-v2). LoRA largely preserves Mandarin accuracy at medium (7.02\% vs. 6.87\%) and large-v2 (5.26\% vs. 5.56\%). This cross-lingual result is consistent with thetrade-off observed in Qwen3-ASR.

The cross-domain degradation measured here characterizes Qwen3-ASR under this six-hour scripted corpus and LoRA recipe, where the pretrained backbone already recognizes Cantonese accurately. Whether other strong backbones show the same trade-off remains to be measured. Part of the CER increase on MagicData-Conv reflects orthographic alignment: conversational references write \raisebox{-0.28em}{\includegraphics[height=1.15em]{figs/zh/z5208de1d00.pdf}} where the competition corpus writes \raisebox{-0.28em}{\includegraphics[height=1.15em]{figs/zh/z0c9cd6e291.pdf}}, and fine-tuned models adopt the training convention. In a normalization sensitivity check on two Qwen3-ASR-1.7B runs at $10^{-4}$ (Section~\ref{sec:appb}), normalizing numerals and folding 15 written variants reduced the excess conversational CER over zero-shot by roughly 54\% (seed 42) and 45\% (seed 43), yet both fine-tuned runs remained substantially worse than zero-shot under the same rules.

\section{Conclusion}\label{sec:conclusion}

We presented W500, a reproducible low-resource Cantonese ASR competition system, alongside an empirical study of cross-domain generalization across weak- and strong-prior pretrained models. W500 implements source-conditional two-stage fine-tuning of Whisper-small via per-step \texttt{requires\_\allowbreak{}grad} and learning-rate masking inside a single AdamW optimizer, reducing MagicData-Conv CER from 62.69\% zero-shot to 25.04\%. Across our twelve-run Whisper capacity matrix, validation and in-domain test sets rank LoRA differently, capacity returns plateau above medium, microbatch-matched SC lowers Val accuracy at small while reducing In-domain Test CER by 9.5\% and 6.0\% at large-v2, and cross-script evaluation requires symmetric normalization. In a matched three-seed control with MDCC external audio, freezing the decoder on external steps costs validation accuracy and does not lower conversational CER relative to a conventional mixed-batch recipe at equal external exposure. Controlled experiments with Qwen3-ASR demonstrate that 6.3 hours of scripted fine-tuning yield strong in-domain gains at a systematic cost to independent spontaneous speech and the public WSYue-ASR-eval benchmark. Lowering the LoRA learning rate tenfold retains in-domain accuracy while cutting conversational degradation by roughly 47\% at 1.7B and 78\% at 0.6B, and the regression persists after normalizing numerals and written variants. Reporting a zero-shot baseline on an independent corpus alongside competition scores provides a direct check to detect such cross-domain trade-offs early.

\FloatBarrier
\bibliographystyle{unsrtnat}
\bibliography{refs}

@misc{dimsum,
  author = {AI Dimsum Cup organisers},
  title = {{AI Dimsum Cup, preliminary Cantonese ASR task (task page)}},
  year = {2026},
  howpublished = {competition page},
  url = {https://www.aicompetition-pz.com/topic_detail/19},
  note = {Accessed 2026-11-30}
}

@misc{lifescen,
  author = {leeduckgo},
  title = {{Cantonese Life Scenarios Corpus (Hugging Face dataset)}},
  year = {2026},
  howpublished = {dataset},
  url = {https://huggingface.co/datasets/leeduckgo/cantonese-life-scenarios-corpus},
  note = {Accessed 2026-11-30}
}

@misc{hankrepo,
  author = {Vanxun-Hank},
  title = {{Cantonese ASR with Whisper-small}},
  year = {2026},
  howpublished = {GitHub repository, commit f2c576f},
  url = {https://github.com/Vanxun-Hank/cantonese-asr/tree/f2c576f7da5d376219bea4246641730713a762a1},
  note = {Accessed 2026-11-30}
}

@misc{hankmodel,
  author = {cantonese-asr-lab},
  title = {{whisper-small-cantonese-w500-adaptive (released model, gated access)}},
  year = {2026},
  howpublished = {Hugging Face model card},
  url = {https://huggingface.co/cantonese-asr-lab/whisper-small-cantonese-w500-adaptive},
  note = {Accessed 2026-11-30}
}

@misc{whisper,
  author = {Radford, A. and others},
  title = {{Robust Speech Recognition via Large-Scale Weak Supervision}},
  year = {2022},
  howpublished = {arXiv:2212.04356},
  url = {https://arxiv.org/abs/2212.04356},
  note = {Accessed 2026-11-30}
}

@misc{wsyue,
  author = {Li, L. and Guo, Z. and Chen, H. and Dai, Y. and others},
  title = {{WenetSpeech-Yue: A Large-scale Cantonese Speech Corpus with Multi-dimensional Annotation}},
  year = {2025},
  howpublished = {arXiv:2509.03959},
  url = {https://arxiv.org/abs/2509.03959},
  note = {Accessed 2026-11-30}
}

@misc{lora,
  author = {Hu, E. J. and others},
  title = {{LoRA: Low-Rank Adaptation of Large Language Models}},
  year = {2021},
  howpublished = {arXiv:2106.09685},
  url = {https://arxiv.org/abs/2106.09685},
  note = {Accessed 2026-11-30}
}

@misc{qwen3asr,
  author = {Shi, X. and Wang, X. and Guo, Z. and others},
  title = {{Qwen3-ASR Technical Report}},
  year = {2026},
  howpublished = {arXiv:2601.21337},
  url = {https://arxiv.org/abs/2601.21337},
  note = {Accessed 2026-11-30}
}

@misc{chinavoices,
  author = {Liao, Y. and Mu, B. and Wang, S. and others},
  title = {{Summary of the ChinaVoices Challenge 2026: Data, Tasks, Baseline, and Methods}},
  year = {2026},
  howpublished = {arXiv:2609.03471},
  url = {https://arxiv.org/abs/2609.03471},
  note = {Accessed 2026-11-30}
}

@misc{magic,
  author = {{Beijing Magic Data Technology Co., Ltd.}},
  title = {{ASR-CCantCSC: A Chinese Cantonese (Canton) Conversational Speech Corpus}},
  year = {n.d.},
  howpublished = {MagicHub dataset},
  url = {https://magichub.com/datasets/guangzhou-cantonese-conversational-speech-corpus/},
  note = {Accessed 2026-11-30}
}

@misc{narrowing,
  author = {Gunduz, K. G.},
  title = {{Winning the Leaderboard, Losing the Distribution: Diagnosing and Mitigating Acoustic Narrowing in Low-Resource ASR}},
  year = {2026},
  howpublished = {arXiv:2611.18653},
  url = {https://arxiv.org/abs/2611.18653},
  note = {Accessed 2026-11-30}
}

@misc{ulmfit,
  author = {Howard, J. and Ruder, S.},
  title = {{Universal Language Model Fine-tuning for Text Classification}},
  year = {2018},
  howpublished = {arXiv:1801.06146 (ACL 2018)},
  url = {https://arxiv.org/abs/1801.06146},
  note = {Accessed 2026-11-30}
}

@misc{surgical,
  author = {Lee, Y. and Chen, A. S. and Tajwar, F. and Kumar, A. and Yao, H. and Liang, P. and Finn, C.},
  title = {{Surgical Fine-Tuning Improves Adaptation to Distribution Shifts}},
  year = {2022},
  howpublished = {arXiv:2210.11466 (ICLR 2023)},
  url = {https://arxiv.org/abs/2210.11466},
  note = {Accessed 2026-11-30}
}

@misc{lorafg,
  author = {Biderman, D. and others},
  title = {{LoRA Learns Less and Forgets Less}},
  year = {2024},
  howpublished = {arXiv:2405.09673},
  url = {https://arxiv.org/abs/2405.09673},
  note = {Accessed 2026-11-30}
}

@misc{bell2020,
  author = {Bell, P. and Fainberg, J. and Klejch, O. and Li, J. and Renals, S. and Swietojanski, P.},
  title = {{Adaptation Algorithms for Neural Network-Based Speech Recognition: An Overview}},
  year = {2020},
  howpublished = {arXiv:2008.06580},
  url = {https://arxiv.org/abs/2008.06580},
  note = {Accessed 2026-11-30}
}

@misc{rouditchenko,
  author = {Rouditchenko, A. and Khurana, S. and Thomas, S. and Feris, R. and Karlinsky, L. and Kuehne, H. and Harwath, D. and Kingsbury, B. and Glass, J.},
  title = {{Comparison of Multilingual Self-Supervised and Weakly-Supervised Speech Pre-Training for Adaptation to Unseen Languages}},
  year = {2023},
  howpublished = {arXiv:2305.12606 (Interspeech 2023)},
  url = {https://arxiv.org/abs/2305.12606},
  note = {Accessed 2026-11-30}
}

@misc{field23,
  author = {Field, A. and Verma, P. and San, N. and Eberhardt, J. L. and Jurafsky, D.},
  title = {{Developing Speech Processing Pipelines for Police Accountability}},
  year = {2023},
  howpublished = {Proceedings of Interspeech 2023, pp. 1229--1233, doi:10.21437/Interspeech.2023-2109},
  url = {https://doi.org/10.21437/Interspeech.2023-2109},
  note = {Accessed 2026-11-30}
}

@misc{du18,
  author = {Du, Y. and Czarnecki, W. M. and Jayakumar, S. M. and Farajtabar, M. and Pascanu, R. and Lakshminarayanan, B.},
  title = {{Adapting Auxiliary Losses Using Gradient Similarity}},
  year = {2018},
  howpublished = {arXiv:1812.02224},
  url = {https://arxiv.org/abs/1812.02224},
  note = {Accessed 2026-11-30}
}

@misc{pcgrad,
  author = {Yu, T. and Kumar, S. and Gupta, A. and Levine, S. and Hausman, K. and Finn, C.},
  title = {{Gradient Surgery for Multi-Task Learning}},
  year = {2020},
  howpublished = {arXiv:2001.06782},
  url = {https://arxiv.org/abs/2001.06782},
  note = {Accessed 2026-11-30}
}

@misc{l2sp,
  author = {Li, X. and Grandvalet, Y. and Davoine, F.},
  title = {{Explicit Inductive Bias for Transfer Learning with Convolutional Networks}},
  year = {2018},
  howpublished = {arXiv:1802.01483 (ICML 2018)},
  url = {https://arxiv.org/abs/1802.01483},
  note = {Accessed 2026-11-30}
}

@misc{wiseft,
  author = {Wortsman, M. and others},
  title = {{Robust fine-tuning of zero-shot models}},
  year = {2021},
  howpublished = {arXiv:2109.01903 (CVPR 2022)},
  url = {https://arxiv.org/abs/2109.01903},
  note = {Accessed 2026-11-30}
}

@misc{mdcc,
  author = {Yu, T. and Frieske, R. and Xu, P. and Cahyawijaya, S. and Yiu, C. T. S. and others},
  title = {{Automatic Speech Recognition Datasets in Cantonese: A Survey and New Dataset}},
  year = {2022},
  howpublished = {arXiv:2201.02419},
  url = {https://arxiv.org/abs/2201.02419},
  note = {Accessed 2026-11-30}
}

@misc{loraint8,
  author = {Zhang, L. and Wu, S. and Wang, Z.},
  title = {{LoRA-INT8 Whisper: A Low-Cost Cantonese Speech Recognition Framework for Edge Devices}},
  year = {2025},
  howpublished = {Sensors 25(17):5404},
  url = {https://doi.org/10.3390/s25175404},
  note = {Accessed 2026-11-30}
}

@misc{koelony,
  author = {koelony},
  title = {{Qwen3-ASR-1.7B-Cantonese-HK (model card)}},
  year = {2026},
  howpublished = {Hugging Face model card},
  url = {https://huggingface.co/koelony/Qwen3-ASR-1.7B-Cantonese-HK},
  note = {Accessed 2026-11-30}
}

@misc{crd,
  author = {Wang, Y. and Zhou, W. and Zhang, Y. and others},
  title = {{Collapse, Repair, and Drift: Dissecting Fine Tuning Dynamics in Speech Language Models}},
  year = {2026},
  howpublished = {arXiv:2610.18192},
  url = {https://arxiv.org/abs/2610.18192},
  note = {Accessed 2026-11-30}
}

@misc{opsd,
  author = {Wang, S. and Mu, B. and Zhang, P. and others},
  title = {{On-Policy Self-Distillation for Multi-Dialect ASR: Mastering Dialects, Retaining Mandarin}},
  year = {2026},
  howpublished = {arXiv:2608.11898},
  url = {https://arxiv.org/abs/2608.11898},
  note = {Accessed 2026-11-30}
}

@misc{cv,
  author = {Ardila, R. and others},
  title = {{Common Voice: A Massively-Multilingual Speech Corpus}},
  year = {2019},
  howpublished = {arXiv:1912.06670 (LREC 2020)},
  url = {https://arxiv.org/abs/1912.06670},
  note = {Accessed 2026-11-30}
}

@misc{aishell,
  author = {Bu, H. and others},
  title = {{AISHELL-1: An Open-Source Mandarin Speech Corpus and A Speech Recognition Baseline}},
  year = {2017},
  howpublished = {arXiv:1709.05522},
  url = {https://arxiv.org/abs/1709.05522},
  note = {Accessed 2026-11-30}
}

@misc{opencc,
  author = {BYVoid and contributors},
  title = {{OpenCC: Open Chinese Convert}},
  year = {2010},
  howpublished = {GitHub},
  url = {https://github.com/BYVoid/OpenCC},
  note = {Accessed 2026-11-30}
}

@misc{cn2an,
  author = {Ailln},
  title = {{cn2an: Chinese and Arabic numeral conversion}},
  year = {2022},
  howpublished = {GitHub},
  url = {https://github.com/Ailln/cn2an},
  note = {Accessed 2026-11-30}
}

@misc{sensevoice,
  author = {An, K. and Chen, Q. and Deng, C. and Du, Z. and Gao, C. and others},
  title = {{FunAudioLLM: Voice Understanding and Generation Foundation Models for Natural Interaction Between Humans and LLMs}},
  year = {2024},
  howpublished = {arXiv:2407.04051},
  url = {https://arxiv.org/abs/2407.04051},
  note = {Accessed 2026-11-30}
}
\appendix

\section{Reproduction}\label{sec:app}

Project code, configurations, verification receipts, and the W500 method note reside in the repository \citep{hankrepo} at commit f2c576f (converged matrix report: \url{https://github.com/Vanxun-Hank/cantonese-asr/blob/f2c576f7da5d376219bea4246641730713a762a1/reports/raw_winner_p2_full_converged.md}), and the released W500 checkpoint is hosted on Hugging Face \citep{hankmodel}. The W500 trainer is \texttt{scripts/train\_\allowbreak{}w500\_\allowbreak{}adaptive\_\allowbreak{}continuation.py} and the SC stream builder is \texttt{scripts/build\_\allowbreak{}source\_\allowbreak{}conditional\_\allowbreak{}stream.py}. The released W500 archive passed flat-ZIP layout checks, SHA-256 verification across weights, model and generation configurations, tokenizer files, and the inference script, and exact transcript and token-sequence matching against the original checkpoint on a fixed 32-utterance offline subset. For this report, \texttt{code/whisper\_\allowbreak{}eval.py} re-scores Whisper checkpoints under transformers 4.57.6 using the repository's D0\_CURRENT preset (2-beam decoding). Qwen3-ASR training and evaluation use \texttt{code/train\_\allowbreak{}qwen3asr.py} and \texttt{code/evaluator.py} (transformers 5.24.1, PEFT 0.22). All numbers in Table~\ref{tab:qwen} and Figure~\ref{fig:fig_qwen_v5} can be recomputed on a CPU via \texttt{code/rescore\_\allowbreak{}qwen.py} from per-utterance prediction files (\texttt{pred\_\allowbreak{}public.jsonl} for In-domain Test, \texttt{pred\_\allowbreak{}ood.jsonl} for OOD Panel, and \texttt{pred\_\allowbreak{}gz\_\allowbreak{}conv.jsonl} filtered by recording IDs A0006, A0018, A0032 for MagicData-Conv). \texttt{code/score\_\allowbreak{}qwen\_\allowbreak{}wsy.py} and \texttt{code/wsyue\_\allowbreak{}score.py} compute Table~\ref{tab:wsyue} from \texttt{pred\_\allowbreak{}wsyue\_\allowbreak{}short.jsonl} and \texttt{pred\_\allowbreak{}wsyue\_\allowbreak{}long.jsonl}. Per-run \texttt{meta.json} files log all training arguments. Because the MagicData corpus is licensed, only utterance ID manifests are redistributed.

Run identifiers. Original 17 Qwen3-ASR runs (Table~\ref{tab:qwen}): zero-shot \texttt{zs\_\allowbreak{}0.6B}, \texttt{zs\_\allowbreak{}1.7B}; challenge recipe \texttt{R1\_\allowbreak{}0.6B\_\allowbreak{}lora\_\allowbreak{}adamw\_\allowbreak{}lr1e-\allowbreak{}4\_\allowbreak{}s{42,\allowbreak{}43}}, \texttt{B2\_\allowbreak{}lora1.7B\_\allowbreak{}adamw\_\allowbreak{}lr1e-\allowbreak{}4\_\allowbreak{}ep3\_\allowbreak{}s{42,\allowbreak{}43}}, and \texttt{R2\_\allowbreak{}1.7B\_\allowbreak{}lora\_\allowbreak{}adamw\_\allowbreak{}lr1e-\allowbreak{}4\_\allowbreak{}s44}; lower learning-rate runs \texttt{R1\_\allowbreak{}0.6B\_\allowbreak{}lora\_\allowbreak{}adamw\_\allowbreak{}lr{3e-\allowbreak{}5,\allowbreak{}1e-\allowbreak{}5}\_\allowbreak{}s{42,\allowbreak{}43}} and \texttt{R2\_\allowbreak{}1.7B\_\allowbreak{}lora\_\allowbreak{}adamw\_\allowbreak{}lr{3e-\allowbreak{}5,\allowbreak{}1e-\allowbreak{}5}\_\allowbreak{}s{42,\allowbreak{}43,\allowbreak{}44}}. An earlier run, \texttt{B2\_\allowbreak{}lora0.6B\_\allowbreak{}adamw\_\allowbreak{}lr1e-\allowbreak{}4\_\allowbreak{}ep3\_\allowbreak{}s{42,\allowbreak{}43}}, produces identical reported scores at therecorded numerical precision (excluded from averaging). Replicated 17 runs on torch 2.10 (Table~\ref{tab:wsyue}): \texttt{QWZ\_\allowbreak{}{0.6B,\allowbreak{}1.7B}\_\allowbreak{}zs} and \texttt{QW\_\allowbreak{}{0.6B,\allowbreak{}1.7B}\_\allowbreak{}lora\_\allowbreak{}adamw\_\allowbreak{}lr{1e-\allowbreak{}4,\allowbreak{}3e-\allowbreak{}5,\allowbreak{}1e-\allowbreak{}5}\_\allowbreak{}s{42,\allowbreak{}43[,\allowbreak{}44]}}. Whisper re-evaluations (Table~\ref{tab:whisper_unified}) use \texttt{W\_\allowbreak{}{small,\allowbreak{}medium,\allowbreak{}large-\allowbreak{}v2}\_\allowbreak{}{full,\allowbreak{}lora}\_\allowbreak{}s{42,\allowbreak{}43}} (\texttt{checkpoint-\allowbreak{}4371}) and \texttt{W\_\allowbreak{}released\_\allowbreak{}w500\_\allowbreak{}small}.

\section{Sensitivity of Conversation CER to Scoring Rules}\label{sec:appb}

To evaluate how sensitive the conversational CER gap is to Cantonese writing conventions, we report a normalization sensitivity analysis over the same three held-out MagicData-Conv recordings (Table~\ref{tab:foldb}). Both scoring protocols remove punctuation and convert predictions and references to simplified Chinese. The additional protocol then normalizes numerals and applies 15 ordered substitutions to both texts (\texttt{code/fold\_\allowbreak{}rules.tsv}; derived from confusions in seed-42 Qwen outputs on the disjoint remaining seven MagicData recordings, excluding the three held-out test recordings). Some substitutions can merge forms with distinct contextual meanings, such as \raisebox{-0.28em}{\includegraphics[height=1.15em]{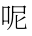}} and \raisebox{-0.28em}{\includegraphics[height=1.15em]{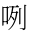}}.

\begin{table}[t]\centering\footnotesize\setlength{\tabcolsep}{3.5pt}\caption{CER (\%) on the three MagicData-Conv recordings under symmetric scoring versus additional numeral and 15-variant normalization. Both fine-tuned Qwen3-ASR-1.7B rows correspond to the B2 LoRA runs at $10^{-4}$ for seeds 42 and 43 in Table~\ref{tab:qwen} (seed 44 was not included).}\label{tab:foldb}\begin{adjustbox}{max width=\linewidth,center}\begin{tabular}{lrrr}\toprule
model & setting & symmetric CER (\%) & with numeral \& variant norm. (\%)\\\midrule
Whisper-small & zero-shot & 62.69 & 62.21\\
Whisper-small & Released W500 & 25.04 & 22.49\\
\midrule
Qwen3-ASR-1.7B & zero-shot & 9.09 & 9.06\\
Qwen3-ASR-1.7B & LoRA, lr $10^{-4}$ (seed 42) & 14.68 & 11.62\\
Qwen3-ASR-1.7B & LoRA, lr $10^{-4}$ (seed 43) & 14.14 & 11.84\\
\bottomrule\end{tabular}\end{adjustbox}\end{table}

Under both protocols, W500 lowers CER substantially relative to zero-shot Whisper-small, whereas both fine-tuned Qwen3-ASR-1.7B runs remain well above zero-shot Qwen (11.62\% and 11.84\% vs. 9.06\%), reducing the excess CER by 54\% (seed 42) and 45\% (seed 43). Qualitative inspection of six utterances drawn from 107 conversation examples (reference length 8--20 characters, at most one edit zero-shot and at least three additional edits after fine-tuning) illustrates both effects. One utterance exhibits equivalent numeral forms (\raisebox{-0.28em}{\includegraphics[height=1.15em]{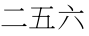}} vs. 256) alongside a genuine digit error (\raisebox{-0.28em}{\includegraphics[height=1.15em]{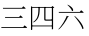}} misrecognized as 326); another shows semantic misrecognition (\raisebox{-0.28em}{\includegraphics[height=1.15em]{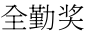}} rendered as \raisebox{-0.28em}{\includegraphics[height=1.15em]{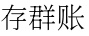}}). Determining what was actually spoken and estimating the exact error breakdown across the full test set would require blinded listening annotations.

\end{document}